# Productive Development of Scalable Network Functions with NFork

Lei Yan, Yueyang Pan, Diyu Zhou, Sanidhya Kashyap, George Candea

EPFL

## Abstract

Despite decades of research, developing correct and scalable concurrent programs is still challenging. Network functions (NFs) are not an exception. This paper presents NFork, a system that helps NF domain experts to *productively* develop concurrent NFs by abstracting away concurrency from developers. The key scheme behind NFork's design is to exploit NF characteristics to overcome the limitations of prior work on concurrency programming. Developers write NFs as sequential programs, and during runtime, NFork performs transparent parallelization by processing packets in different cores. Exploiting NF characteristics, NFork leverages *transactional memory* and develops efficient concurrent data structures to achieve scalability and guarantee the absence of concurrency bugs.

Since NFork manages concurrency, it further provides (i) a profiler that reveals the *root causes* of scalability bottlenecks inherent to the NF's semantics and (ii) actionable recipes for developers to mitigate these root causes by relaxing the NF's semantics. We show that NFs developed with NFork achieve competitive scalability with those in Cisco VPP [16], and NFork's profiler and recipes can effectively aid developers in optimizing NF scalability.

## 1 Introduction

The performance of software network functions (NFs) [20] is essential to today's Internet and data center networks; they conduct critical tasks such as firewalling, NAT, load balancing [14], and mobile core functions [23].

To handle the ever-growing user demands, the line rate of modern servers is increasing to hundreds of Gbps [8, 62]. Meeting such a line rate is challenging for even very simple NFs. Assuming the *smallest* packet size of 64 bytes, to saturate a 100 Gbps network port, an NF must process one packet within 6.7ns on average. Achieving this is impossible with a single core today. As a result, developers exploit multicore architecture by designing multithreaded NFs, as adopted in both industry [14, 16, 40, 42] and academia [12, 24].

Unfortunately, despite decades of research, developing correct (i.e., no concurrency bugs) and scalable (i.e., throughput increases with the number of cores) concurrent programs remains a challenge [34, 52, 55], even for developers with extensive expertise. Perhaps unsurprisingly, NFs are not an exception. We conduct the first study (to the best of our knowledge) on NF concurrency bugs on VPP [16], a mature and widely used networking framework developed by Cisco. We confirm that both correctness and performance concurrency bugs widely exist in VPP NFs, and fixing them is non-trivial, usually requiring multiple iterations and months of effort. Interestingly, VPP developers attempt to avoid concurrency programming by partitioning the NF state among threads to form a shared-nothing architecture. Unexpectedly, such an approach requires (much) more complicated synchronization among threads and thus often, in turn, leads to concurrency bugs.

Existing work is insufficient to enable the productive development of concurrent NFs. Current NF frameworks aid developers in scaling NFs across *different* machines [18, 27, 44, 58] and/or simplifying programming sequential NFs [13, 25, 30, 36, 43]. To our knowledge, none of them targets concurrent NFs on a *single* machine. Generic techniques for easing concurrency programming have limitations when applied to NFs (as detailed in §2.1). Notably, automated loop parallelization techniques heavily depend on the independence of each loop iteration, making them unsuitable for NFs. Concurrency bug detectors [6, 39, 54] do not address the challenge of fixing them. Systems that mask concurrency bugs during execution [31, 32, 35] suffer from high overhead and can only offer (partial) protection against specific types of concurrency bugs. Finally, current profilers cannot reveal the root causes of scalability bottlenecks [10, 11, 52, 57] nor provide effective and actionable suggestions to improve scalability.

This paper presents NFork, an NF programming framework with the associated runtime system and toolchains, to help NF domain experts *productively* develop concurrent NFs without reasoning about concurrency. To achieve this, NFork offers a comprehensive solution that supports programming, profiling, and optimizing NFs. A key insight behind NFork is that the limitations of the aforementioned prior work can be overcome by leveraging specificities of the NF domain, as detailed below.

NFork's programming model offers intuitive abstractions that hide concurrency, which in turn avoids most concurrency-related problems at the source. With NFork, developers write their NFs as sequential programs. This allows them to achieve high productivity, shorter testing cycles, and provide bug fixes and new features to their users and customers more quickly in NFork than otherwise. Unlike prior work on automatic parallelization discussed above, NFork exploits NF characteristics to parallelize the processing of *packets*. To facilitate parallelization, observing that NFs typically classify packets into flows and process packets by accessing state specific to their flows [44], NFork introduces the *packet set* abstraction—a group of logically connected packets—to generalize packet flows.

The NFork runtime parallelizes the sequential NFs provided by the user across available cores transparently, correctly, and efficiently. NFork achieves transparency by parallelizing the processing of different packet sets —each associated with a core—across multiple cores. Instead of following the conventional approach to aid the daunting process of identifying and fixing concurrency bugs, NFork ensures *correct-by-construction* concurrency by revisiting the transactional memory abstraction [21, 35]. Each packet's processing occurs in a transaction. Together with the sequential programming model, this eliminates the possibility for NF developers to introduce concurrency bugs (e.g., data races and atomicity violations). Critically, we find that transactional memory aligns well with NFs; NFork resolves challenges in applying transactional memory (e.g., code regions are too long for a transaction) by exploiting NF characteristics (e.g., packet processing code is short). Finally, NFs often utilize common data structures to manage state, with most concurrent accesses occurring in these structures. Thus, NFork achieves efficiency by offering a set of scalable concurrent data structures. These data structures encapsulate common concurrency wisdom, such as fine-grained synchronization.

With the NFork runtime presented above, NFs parallelized by NFork often achieve good scalability. However, some NFs are inherently difficult to scale due to their semantics (e.g., NFs imposing frequent updates to shared state). To aid developers in productively optimizing these NFs, NFork includes a profiler and a set of scalability improvement *recipes*. The profiler first reports the number of transaction aborts caused by specific lines of code in an NF. This metric acts as an effective proxy for assessing the impact on scalability.

More importantly, unlike existing profilers, the NFork profiler reveals the *root causes* of the scalability bottlenecks. This is possible since NFork manages concurrency, allowing it to bridge the semantic gaps by informing the profiler of a set of *conflict causes*. The conflict causes encode the exact reasons for transaction aborts in built-in concurrent data structures. For each conflict cause, NFork provides *recipes* for improving scalability by relaxing the NF's semantics (e.g., allowing for reading of a stale value). Developers can readily follow these recipes without understanding the intricacy of concurrency.

Our evaluation shows that NFs developed with NFork achieve scalability on par with functionally equivalent but manually parallelized and heavily optimized (including the use of various lock-free algorithms and hardware-specific optimizations like prefetching) NFs from Cisco VPP [16]. We also present three case studies to show how, under the guidance of NFork, one can productively identify and remove the root causes of scalability bottlenecks in NFs by relaxing semantics, thereby improving NF throughput from 2× to 91×.

In summary, we make the following contributions:
- **NFork.** By exploiting NF characteristics, NFork frees NF domain experts from reasoning about concurrency yet allows them to *productively* develop NFs that are free of concurrency bugs and have competitive multicore performance.
- **Profiler and recipes.** The NFork scalability profiler and recipes can quantitatively guide developers in trading the NF's semantics for scalability, thereby improving their productivity in optimizing their NFs atop NFork.
- **NF concurrency bug study.** We conduct the first study (to the best of our knowledge) of NF concurrency bugs with several key findings unique to the NF domain.

## 2 Background and Motivation

Developing concurrent software (i.e., software with multiple threads that accesses shared state) is challenging. This section motivates NFork by first discussing state-of-the-art concurrency programming techniques and explaining why most are unsuitable for NFs (§2.1). Next, we discuss the state of practice in developing multithreaded NFs (§2.2), followed by our study on concurrency issues in NFs (§2.3). We find that, while some bug characteristics are similar to those in generic server applications, there also exist several NF-specific ones.

### 2.1 Related work

**NF programming frameworks.** Prior NF programming frameworks provide abstractions to hide low-level details from programmers [25, 30, 36, 43]. For example, mOS [25] provides abstractions to hide flow processing. There are NF frameworks that help scale NFs across multiple machines in a distributed system setup [18, 27, 44, 58]. However, to the best of our knowledge, none of the prior work aims to help programmers develop multithreaded single-machine shared-state NFs.

**Automatic parallelization.** Work that converts sequentially accessed data structures into concurrent ones [9, 48, 59] heavily depends on the semantics of the data structures and does not readily carry over to NFs. Systems that parallelize general sequential programs focus on automatically extracting parallelism in loops and/or function calls [4, 19, 53]. These parallelization techniques heavily depend on the independence of each loop iteration or function call, making them unsuitable for many NFs (e.g., stateful NFs, where the processing of one packet depends on state changes resulting from processing previous packets).

**Concurrency bug detection and masking.** There has been a long line of work that aids programmers in detecting concurrency bugs [6, 39, 54]. However, even with effective detection, fixing concurrency bugs is still challenging [34]. Existing work on masking concurrency bugs during program execution only targets a specific type of bugs and introduces non-trivial overhead, while not guaranteeing the absence of bugs [31, 32, 35].

**Identifying and fixing scalability bottlenecks.** Besides correctness bugs, concurrency programming is also prone to performance bugs by introducing scalability bottlenecks. Existing profilers [10, 11, 57] identify the *symptoms* of scalability bottlenecks but not the *root cause*, which is necessary for a correct fix. For example, profilers can reveal symptoms such as high contention on critical sections. However, such a symptom may be due to incorrect use of synchronization mechanisms (e.g., use exclusive locks instead of readers-writer locks for read-mostly data) or improper synchronization granularity (e.g., coarse-grained locks to protect all shared state). Existing profilers cannot tell which one is the root cause. Some recent work aims at providing guidelines for fixing scalability bottlenecks [52,60], but it is often too generic (e.g., "use fine-grained locks") to be actionable for developers.

| Bug type | Number of bugs |
| --- | --- |
| Data race | 16 |
| Atomicity violation | 8 |
| Deadlock | 4 |
| **Total** | **28** |

Table 1: Concurrency bugs in VPP NFs.

### 2.2 Current practice of developing concurrent NFs

We study the current practice of concurrent NF development by analyzing the source code, bug database, and mailing list in commercial networking frameworks, such as Cisco VPP [56]. We find that, due to the lack of programming frameworks and the limitations of existing techniques, programmers nowadays still use low-level synchronization primitives, such as atomic instructions and locks, to develop concurrent NFs. They may use certain tools (e.g., perf) to help identify concurrency bugs or scalability bottlenecks but often fix them through trial and error. Such an approach is error-prone, allows bugs to escape, and is highly unproductive, as evidenced below.

### 2.3 NF concurrency problems

**Methodology.** We use concurrent NFs in VPP for our study. VPP is a mature (developed over seven years) and widely used networking framework (shipped over $1 billion of Cisco products [56]). As with prior work [34], we find NF concurrency bugs by first searching concurrency-related keywords (e.g., "deadlock", "atomicity") in the bug database. We then manually analyze randomly chosen bugs and confirm that these are indeed concurrency bugs manifested in NFs.

We summarize our findings below. The first two findings match concurrency bug studies in generic server applications [34], while the last two are specific to NFs.

**Finding #1: Concurrency bugs are prevalent and harmful.** Table 1 summarizes the concurrency bugs we find. Out of the 30 concurrent NFs in VPP, our incomplete and random bug search finds that at least 14 NFs once have concurrency bugs, indicating that concurrency bugs widely exist in NFs. Furthermore, all of the found bugs are harmful: they cause the NF to produce incorrect output, crash, or hang.

**Finding #2: Debugging concurrency bugs in NFs remains challenging.** Identifying and fixing concurrency bugs remains challenging. For the concurrency bugs we found, the time between introducing and finding the concurrency bugs is, on average, 27 months, with the shortest one being 11 months and the longest one being 75 months.

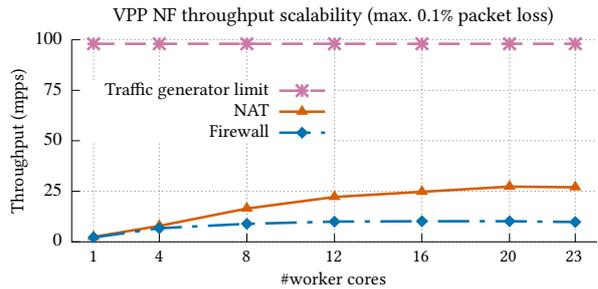

Figure 1: Throughput scalability of VPP NAT and Firewall

In addition, fixing concurrency bugs is non-trivial and often requires multiple iterations. Only one of the studied concurrency bugs is fixed with 1 patch, while others require 2 – 7 patches with an average of 3.1 patches per fix.

**Finding #3: Shared-nothing architecture is not a panacea.** To avoid the complexity of concurrency, a common design in NFs is to employ a shared-nothing architecture, where the NF state is partitioned among threads. Ironically, due to the nature of NFs, a shared-nothing design may still have concurrency bugs, as detailed below.

One bug [1] we find is on NAT's hairpinning mode, where the address mappings are partitioned among cores, and each thread owns a part of the mapping. The bug is that when a thread processes a packet, it reads the address mapping of the packet's destination endpoint without synchronizing the owner thread. This leads to race conditions, e.g., the owner thread frees the address mapping when another thread is still accessing it.

The bug is fixed by checking and forwarding the packet to the thread that owns the address mapping of the packet's destination endpoint. This fix is the most complicated one in our study, resulting in 877 lines of changed code. Passing packets between threads to achieve correct shared-nothing is the common design in the VPP NAT across both normal and hairpinning modes. Unfortunately, as detailed below, it causes a scalability bottleneck.

**Finding #4: Scalability bottlenecks are caused by simple concurrency patterns.** To study the scalability, we configure and run four VPP NFs. Out of the four NFs, two of them do not scale and are far from reaching the 98 mpps limit of our traffic generator with one NIC port, as shown in Figure 1. Our analysis finds that, for NAT, the scalability bottleneck lies in contention on queues that forward packets between cores. For Firewall, the scalability bottleneck is caused by an allocator that inefficiently manages session table entries.

**Conclusion.** We believe that VPP developers, at least some of them, have extensive concurrency expertise; we have seen complex concurrency-related optimizations across many VPP NFs, such as the use of various lock-free algorithms.

However, developing concurrent software is fundamentally challenging [34]. Developing concurrent code requires *non-intuitive*, *non-local* reasoning about the interaction among threads. Concurrency bugs and performance bottlenecks are often non-deterministic, i.e., they can only be triggered by certain inputs. Thus, even with concurrency expertise, developers are prone to introduce concurrency problems and find it difficult to fix them, as evidenced above.

## 3 Target users and NF models

A key insight behind NFork is that, by exploiting the unique properties in NFs, it is possible to design an NF framework that can automatically, efficiently, and correctly parallelize NFs, thereby overcoming the limitations of prior work (§2.1). This section presents the target audience of NFork (§3.1) and a common NF model that NFork builds on (§3.2).

### 3.1 Target audience

NFork targets developers with NF domain expertise. We assume that the developers understand the NF at the semantic level; despite the subtleties of network protocols, given certain incoming packets, the developers understand what are the correct output and state changes of an NF. Furthermore, given the NF's semantics, the developers can implement such an NF, possibly using frameworks such as DPDK [13]. The developers have knowledge useful to optimizing NF performance, and particularly whether trading some aspect of NF's semantics for increased performance makes sense (e.g., they know whether it is acceptable for an NF to use certain stale data if that improves scalability).

NFork aims to help such NF domain experts with or without concurrency expertise. For the domain experts without concurrency expertise, i.e., they are not comfortable working with low-level synchronization primitives nor understanding high-level synchronization algorithms, NFork enables them to develop functioning concurrent NFs with competitive multicore performance. For domain experts with concurrency expertise, NFork helps quickly provide a reasonable implementation. Only for the rare cases where NFork cannot provide satisfactory performance, the developers need to go through the laborious process of developing concurrent NFs.

### 3.2 Target NF model

Our model assumes that an NF operates in three stages: initialization, normal operation, and termination. In the initialization and termination stages, an NF creates or destroys its state by allocating and freeing the corresponding resources, respectively. During normal operation, an NF runs in an event loop. In each iteration, it refers to the NF state to decide how to process a received packet and whether to send out an output packet. An NF may modify, drop, or simply preserve the packet, or generate a new packet as the output packet. The NF also updates its state according to the processing decision.

In addition, NFs typically have the following properties:

**P1: Each flow executes in (mostly) isolated context.** As noted in prior work [44], often an NF classifies a group of logically connected packets (e.g., packets belonging to the same TCP stream) into a "flow". When processing a packet, in most cases, an NF only needs to read or modify the state that is *exclusively* associated with the flow of the packet. State can also be shared among flows, but often it is small and infrequently accessed.

**P2: Short packet processing time.** While a generic server application may take arbitrarily long to produce a response, NFs are generally highly optimized to process packets within a short amount of time (ranging from nanoseconds to, at most, microseconds), since delays in intermediate network nodes can significantly affect end-to-end networking performance and responsiveness.

**P3: No ordering requirement on processing packets in different flows.** An NF may require a specific packet processing order within a *single* flow. For example, packets in a TCP stream may need to be processed in their respective order. However, an NF typically does not require an order between processing packets in two *different* flows. While different orders may result in different outputs, they are all valid.

**P4: Concurrent accesses mostly in a small set of widely used data structures.** NFs typically maintain their state with widely used data structures [61] (such as arrays, key-value maps, allocators, and counters). Furthermore, most of the NF concurrent accesses occur in these data structures, making their scalability critical to the overall NF scalability.

**P5: Restricted interactions with the underlying systems software.** Systems software that supports generic applications needs to provide a large number of functionalities. As evidence, the Linux kernel and the glibc library provide hundreds of systems or library calls. However, as discussed later, NFs only interact with the underlying systems software in a limited way, requiring only a few commonly used functionalities.

We analyze the source code and descriptions of 14 widely used NFs (Table 6) and confirm that they (mostly) match the model discussed above. 11 out of the 14 NFs are organized around a certain type of flows. We do not find the use of ordering synchronization primitives (e.g., conditional variables and barriers) in the NF code, nor do we believe there is a need for doing so according to the NF's semantics. The reported average packet processing time typically varies from 90ns to $3\mu$s, with one exception being the web cache, which takes $50\mu$s. We also found that these NFs perform all concurrent accesses in the aforementioned data structures. Finally, these NFs only require from the underlying system software dynamic memory management, reading the current time, and disk/networking I/O.

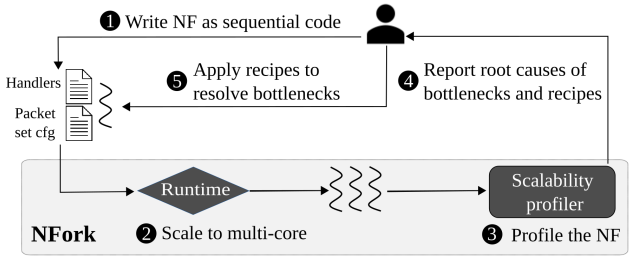

Figure 2: Workflow of NFork

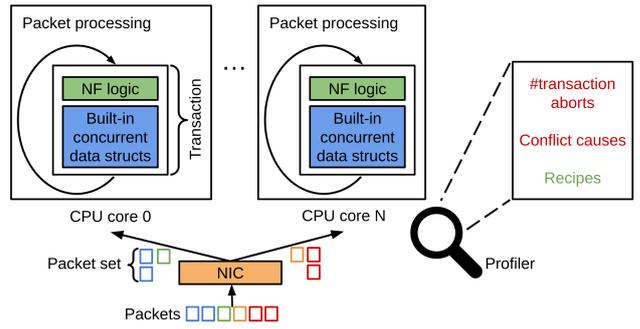

Figure 3: NFork system architecture. Packets of the same color are logically related and belong to the same packet set.

| Handlers | When to invoke |
| --- | --- |
| init_NF_handler | NF starts |
| exit_NF_handler | NF terminates |
| init_pkt_set_handler | The first packet of a packet set arrives |
| pkt_handler | A packet of a live packet set arrives |
| expired_pkt_set_handler | The packet set expires |
| orphan_pkt_handler | An orphan packet arrives |
| periodic_handler | The recurring timer fires |

Table 2: Handlers in the NFork programming model.

## 4 The NFork framework

The challenges to develop concurrent NFs (§2.3), even for developers with extensive concurrency expertise, motivate NFork. NFork abstracts away concurrency, thereby addressing most NF concurrency problems at their source.

This section first presents NFork's design goals (§4.1), its overview and workflow (§4.2), followed by the design of each component (§4.3 – §4.6), with a focus on how NFork exploits NF characteristics to overcome design challenges. We conclude the section with an example of developing and optimizing a concurrent NF with NFork (§4.8).

### 4.1 Design goals (and non-goals)

The *overarching* goal of NFork is to help NF domain experts productively develop scalable NFs. To achieve this, we design NFork to meet the following sub-goals:

- **Intuitive abstractions to hide concurrency.** To enable high development velocity without an excessive learning curve, the abstractions that NFork introduces to hide concurrency should be easily understandable with NF domain knowledge.
- **Encapsulated concurrency wisdom for competitive performance.** While we do not aim for concurrent NFs running on NFork to always outperform hand-parallelized ones, we do want NFork to be competitive for a wide range of NFs. To achieve this, NFork should encapsulate general concurrency wisdom in its abstraction to maximize scalability.
- **Correct-by-construction concurrency.** Finding and fixing concurrency bugs is challenging, even for developers with extensive concurrency expertise (§2.3). Instead of providing support to detect and fix concurrency bugs, NFork should *eliminate* the possibility for developers to introduce concurrency bugs. However, preventing bugs that occur in sequential programming (e.g., buffer overflow) is out of scope.
- **Rapid scalability optimization.** It often takes many iterations of identifying and fixing scalability bottlenecks to achieve the desired performance. Thus, NFork should include support to simplify this process.

### 4.2 NFork overview and workflow

Figure 2 shows the workflow and key components of NFork. At its core, NFork exposes an event-driven sequential NF programming model to developers. NFork exposes a set of event handlers (Table 2). ❶ Developers program NFs by writing *sequential* NF code within these handlers (§4.3). ❷ NFork then takes the sequential code provided by the developer and transparently scales it to run on multiple cores with the NFork runtime (§4.4-§4.5). We describe later how we co-design the NFork programming model and the runtime to a) ensure that the programmers cannot introduce concurrency bugs; and b) extract fine-grained parallelism in the NF code to maximize scalability.

Our experience suggests that, in most cases, NFs parallelized by NFork achieve good scalability. However, a user may provide NF code that cannot scale due to inherent heavy contention on shared state. ❸ In this case, the user profiles the NF with the NFork profiler (§4.6). ❹ The profiler directly reveals to the user the scalability bottleneck with an explanation of its root cause. Furthermore, for each possible root cause, NFork provides the developer with one or several actionable *recipes* to remove the scalability bottleneck. The recipes are easy to apply, typically only require changing one or a few lines of NF code, and guide the developer to make well-thought-out and quantitative trade-offs between the NF's semantics and scalability (§4.6). ❺ Thanks to the NFork profiler and recipes, NF developers can productively improve NF scalability (§4.6).

Figure 3 shows the NFork system architecture.

### 4.3 NFork programming model

**Packet set as the unit of parallelism.** A key challenge NFork must overcome is to identify the parallelism in the sequential program for transparent scaling where prior work falls short (§2.1). We resolve this challenge by exploiting the NF property P1 (§3.2) to schedule the processing of different *packets* across multiple cores.

NFork introduces the *packet set* abstraction to generalize the concept of flow (§3.2). A packet set is a group of packets that are logically connected. Leveraging the observation that each flow forms an *execution context* (NF property P1), NFork minimizes synchronization overhead by processing all packets in a packet set on the same core, as further detailed below. Packets that do not belong to any packet set (which we call *orphan packets*) are evenly distributed across cores.

**Programming with NFork.** To use NFork, a programmer first defines packet sets in a config file by specifying the network protocols and the header fields in packets that uniquely map to a packet set (Listing 1). Afterward, the programmer completes a set of event handlers in plain C (Table 2) to implement NF logic. Event handlers that the developer does not implement are no-ops.

We design the handlers to make them generic enough to express most NFs. The first two handlers are used to initialize and clean up NF, respectively. Similarly, for each packet set, there are handlers to initialize it and clean it up. The initialization handler is invoked when the first packet of a packet set arrives. After the initialization, the packet set is *live*, and the `pkt_handler()` is used to process subsequent packets. The cleanup handler is invoked when a packet set expires (i.e., has not received any packet for a certain amount of time specified during initialization.). NFork also provides handlers for processing "orphan packets" and periodic events (e.g., monitoring the health of load balancer backends).

Users can use any statements supported in standard C to program the above handlers. In addition, NFork provides users with a set of library calls to support the basic NF functionalities discussed in §3.2 and common data structures used by the NF (§4.5). However, using any concurrency-related variable types (e.g., atomic) or library calls (e.g., the pthread library) is prohibited to prevent programmers from introducing concurrency bugs, as further discussed below.

**Local vs. global state.** With the packet set abstraction, the NF state is of two kinds: a) *packet-set state*, which is associated *exclusively* with one packet set and thus is *only* accessed by the NF when it processes packets in that packet set; b) *aggregate state*, which is shared across packet sets, and thus may be accessed by the NF when it processes any packet. Since NFork processes all packets in a packet set on the same core, there is no need to synchronize access to the packet-set state. Only access to the aggregate state needs to be synchronized.

To classify these two types of state, NFork currently requires developers to annotate the variable definition with two new types: `pkt_set_state` and `aggregate_state`. In the future, we expect manual annotations to be obviated by tools like [28]. Afterward, the NFork runtime instruments access to the aggregate state to ensure correct and scalable parallelization, as detailed next.

### 4.4 Automatic scaling with NFork runtime

**Efficient and generic concurrency control with transactions.** The mechanism that NFork uses to synchronize access to the aggregate state should achieve good scalability. In addition, it should ensure the concurrent NFs perform intuitively to the developer, i.e., as with the sequential implementation, it processes each packet atomically.

To that end, among several options, such as lock inferences [33] and RCU [37], we choose *transactional memory* [21, 49] as the synchronization mechanism. While transactional memory cannot always achieve the best performance, it can achieve good performance under most scenarios by exploiting fine-grained parallelisms. Specifically, two transactions can execute concurrently as long as there is no conflict access (i.e., concurrent read-write or write-write access) to the same object. More importantly, the behavior of transactional memory is also simple and well-understood; in brief, it ensures that concurrent instances appear to execute sequentially and atomically (i.e., all-or-nothing semantics for operations within each instance). We defer the discussion on how NFork ensures atomicity in the presence of events, such as I/Os, towards the end of the section.

Following this design choice, NFork executes each of its handlers in a transaction; the transaction starts at the beginning of a handler and commits when the handler returns. During the execution of the handler, NFork tracks the accesses to aggregate state so that it can abort the transaction if there is a conflict access to the aggregate state. In such a case, NFork retries the aborted transactions. (We discuss liveness in §6). The NFork transaction is implemented in software based on MVRLU [29].

**Avoiding concurrency bugs.** With NFork, the sequential programming model paired with transactional access to shared state eliminates concurrency bugs by design. Considering the common type of concurrency bugs, NFork eliminates deadlocks and order violations since no locks or ordering primitives are required (P3 in §3.2) or allowed in NF code. Both data races and atomicity violations are masked by the transactional memory. Upon manifestation, these two types of bugs lead to a transaction abort. Thus, once NFork re-executes and successfully commits the transaction, these two types of bugs are guaranteed to have not manifested. As shown in Table 1, all of the 28 bugs we found in VPP belong to these types.

**Overcoming challenges in applying transactional memory.** Transactional memory simplifies concurrency programming [21, 35] but faces many unexpected challenges in real-world applications [34, 55]. We describe how NFork addresses these challenges below.

Synchronization mechanisms and I/O operations are ill-suited to transactional memory. The former avoids conflicting accesses, which is pointless given that a transaction automatically aborts on such conflicts. The latter requires com-plicated mechanisms for undoing its effects upon transaction aborts. To resolve this issue, our programming model prevents the use of synchronization primitives and uses buffering I/O; all the I/O requests in transactions are buffered in memory and are discarded or performed when the transaction aborts or commits, respectively. NFork ensures that all the code of an NF handler is and only in one transaction, because nested transactions and the interaction between code in and outside transactions leads to complexity and non-intuitive semantics.

Since NFork's transactional memory is based on automatically instrumenting NF accesses to aggregate state, executing code outside NF (e.g., library and system calls) causes a problem; the NFork runtime is not able to track and thus revert such state changes (e.g., file descriptor state changes due to `seek()`). Fortunately, unlike generic applications, NFs only require a few features from the underlying system software (P5 in §3.2). NFork thus implements these functionalities in its runtime and integrates them with transactional memory. Finally, generic applications may have long code regions that cannot be wrapped in a transaction, while NFs typically do not face such constraints (P2 in §3.2).

## 4.5 Efficient scaling with built-in data structures

NFork provides a set of built-in concurrent data structures, to leverage the fact that most concurrent accesses in an NF go to a small set of common data structures (P4 in §3.2). NFork embeds concurrency expertise in them by employing advanced optimizations, including pre-allocation and replication.

Developers may use one or more built-in data structures within a handler by invoking the corresponding APIs. Since built-in data structures are called from a handler, transactional memory ensures safe access. Within a handler (and thus within a transaction), operations across different data structures execute atomically. Conflict accesses performed by built-in data structures abort the transaction. We next detail each of the built-in data structures, namely vector, allocator, map, and distributed object, focusing on their interfaces and the enhancements to maximize scalability. The library of built-in data structures can be easily extended by concurrency experts.

**Vector.** The vector is an array of elements with `read()` and `write()` interfaces. In the vector, each element is aligned with the cache line size to prevent false sharing. To minimize memory overhead, we keep memory-transaction metadata in the spare space created by the alignment constraints. The same alignment approach is used for other data structures as well.

**Allocator.** NFs manage resources (e.g., public IPs) using the allocator. To meet common semantics in NFs, the allocator associates an expiration time to each allocated object. The allocator frees an object after its time expires. In addition to the usual `allocate()` and `free()` interfaces, the resource allocator provides a `refresh()` function that enables the owner of an allocated object to reset its expiration time.

| | allocate() | | refresh(res2) |
|---|---|---|---|
| allocate() | Conflict causes:<br>(1) A local pool becomes exhausted, causing contention on another pool.<br><br>Recipe:<br>(1) Overprovision resource. | refresh(res1) | Conflict causes:<br>(1) res1 and res2 placed on the same allocated list.<br><br>Recipe:<br>(1) Reduce refresh interval. |

**Table 3:** Conflict causes and recipes of allocator. We omit some due to space limitations.

To maximize scalability, our design makes each core maintain a local pool of free resources. When its local pool becomes exhausted, a core obtains free resources from the core with the most free resources. To facilitate object reclaim, each core maintains a list of allocated objects, sorted by the remaining expiration time in descending order. Each `refresh()` moves the object to the list head.

**Map.** The map maintains a one-to-one mapping between a key and a value, useful for data structures like the MAC table in a bridge NF. The map provides a typical `get()` and `set()` interface. We implement the map as a bucket-based hash table to maximize scalability. Thus, concurrent accesses to different buckets proceed in parallel, and only conflict accesses within the same bucket abort the transaction.

**Distributed object.** We design the distributed object for objects with concurrent write- and read-write-intensive accesses (e.g., global counters in NF). A distributed object has `read()` and `update()` interfaces. The distributed object stores a portion of the object in each core. `Update()` directly modifies the local portion of the invoking core, thereby avoiding the contention. `Read()` requires iterating and merging the local portion of each core with a merging function, that is provided by the NF developer during the initialization of the distributed object, to obtain the final result. To mitigate the read-write contention, we make the distributed object store the last merged value, the current staleness (i.e., the time elapsed since the last merging), and the maximal staleness (which is specified by the NF developer during the initialization of the distributed object). `Read()` does not iterate but directly returns the last returned value if the object's staleness is smaller than the maximal staleness. Thus, developers can reduce read-write contention by increasing the maximal staleness.

## 4.6 Using the profiler and recipes to improve NF scalability

Existing profilers rarely reveal directly the root causes of scalability bottlenecks nor do they provide effective guidelines to improve scalability (§2.1). Such inefficiency lies in the *semantic gap* between what the profiler sees and what the application does. The profiler can, at best, understand the semantics of low-level synchronization primitives (e.g., locks and atomic operations); it cannot understand the high-level concurrent data structures and algorithms, thereby failing to identify the root cause behind the scalability issue.

| *Map* | `get(key2)` | `set(key2)` |
|---|---|---|
| `get(key1)` | No conflict | Conflict causes:<br>(1) key1 equals key2.<br>(2) Two keys hashed to the same bucket.<br><br>Recipes:<br>(1) Use distributed objects.<br>(2) Increase map size or change hash func. |
| `set(key1)` | | Same as `get(key1)` and `put(key2)` |

**Table 4:** Conflict causes and recipes of map. The diagonal line means the operation pair is already presented in another cell.

| *Distributed object* | `read()` | `update()` |
|---|---|---|
| `read()` | No conflict | Conflict causes:<br>(1) Object exceeds maximal staleness (thus read merges local portions).<br><br>Recipe:<br>(1) Increase maximal allowed staleness. |
| `update()` | | No conflict |

**Table 5:** Conflict causes and recipes of distributed objects. The diagonal line has the same meaning as in Table 4.

Since NFork completely manages the NF's concurrency, it can bridge this semantic gap. It overcomes the challenge by having the runtime, specifically the built-in concurrent data structures (§4.5), to convey key knowledge, such as the intended behavior of each interface, to the NFork profiler. Therefore, the profiler directly distills to developers the root causes and the correct fixes of the scalability bottlenecks.

**Bridging the semantic gap with conflict causes.** NFork provides the *conflict causes* abstraction to bridge the profiler with the built-in data structures. Each built-in data structure includes a set of conflict causes. A conflict cause specifies the root cause of a transaction abort. It takes the form of a condition between a pair of data-structure operations that, if met, leads to conflicting concurrent accesses that will abort the transaction.

**Using recipes to productively improve scalability.** Each conflict cause further comes with one or more *recipes* to guide developers on mitigating or eliminating the associated scalability bottleneck. The recipes encapsulate what a concurrency expert would do when encountering the same scalability bottleneck. Examples of the recipe include overprovisioning resources, allowing for stale reads, or employing other suitable built-in data structures.

Conflict causes provide NFork users with the precise reason for a scalability bottleneck, and the associated recipes provide solutions for optimizing the bottleneck. Tables 3- 5 show the conflict causes and recipes of the allocator, map, and distributed object, respectively. For vector, the conflict causes are concurrent read-write or write-write to the same element. The recipes are to replace contended elements with distributed objects.

The developer of the built-in concurrent data structures specifies its conflict causes and the recipes. We derive the conflict causes and recipes of each built-in data structure discussed above by considering their implementations. As an illustrative example, since the map is a bucket-based hash table (§4.5), one conflict cause is that two `set()` contend on the same bucket. The recipe is thus either to increase the map size or use a better hash function for load balancing to avoid such contention.

We design the interfaces of built-in data structures so that most recipes can be applied by simply changing the parameters in the initialization function. Recipes (e.g., overprovisioning resources) often involve relaxing the NF's semantics, i.e., allowing additional behavior other than those specified in the specification, exposing a trade-off between NF scalability and semantics. With NFork, developers can use their domain knowledge to identify the sweet spot.

**NFork scalability profiler.** The profiler reports two kinds of information. The first one is the number of transaction aborts (and the fraction of total transaction aborts they represent) caused by each line of the NF code. The number of transaction aborts serves as a "proxy" metric for the scalability cost of shared state contention. The profiler reports the second type of information when the transaction abort occurs in a built-in data structure. In this case, the profiler first collects which conflict causes are responsible for the abort. The profiler then ranks conflict causes based on the number of transaction aborts they cause and reports them. For each conflict cause, the profiler also reports the corresponding recipe(s).

### 4.7 Implementation

We implement NFork as a userspace C library on top of the DPDK [13] kernel bypass, which NFork uses for network I/O. Inside each transaction, to minimize overhead while correctly detecting conflict accesses, NFork tracks memory accesses at the object granularity (instead of, say, at the byte granularity).

**Distributing packets with RSS** To minimize synchronization overhead, NFork distributes packet sets among cores and achieves load balancing by leveraging the NIC RSS [46]. RSS distributes packets among NIC receive queues, which are pinned 1:1 to the CPU cores, based on the value of a set of packet header fields. By configuring RSS to use all or a subset of the packet header fields specified in the packet set definition, NFork guarantees that all packets of a packet set are directed by RSS to the same core.

NFork includes an automated checker that tells developers whether their packet set definitions are compatible with the NIC's RSS. The RSS checker is meant to be part of the NF developers' toolchain, to be used both during design and during compilation. In the case of incompatibility, the checker reports the unsupported packet header fields. Developers can then choose to not define packet sets or try to change the

```
/* Each TCP flow is a packet set.
 * NIC 0 receives the packets. */
[{nic: 0,
  pattern: {"ipv4":    [src_ip, dst_ip],
            "tcp": [src_port, dst_port]}
}]
```
Listing 1: Packet set definition of Anti-DDoS.

definition to avoid the unsupported header fields. However, in-compatibility happens rarely and the packet set definitions of all NFs in Table 6 are supported by mainstream 100Gbps NICs like Mellanox Connectx5 and Intel E810.

**Transaction batching.** Many NFs process each packet fast (the fastest one is 90ns in §5.3). In these cases, NFork batches the NF handlers of multiple packets in a single transaction to amortize the overhead of starting and committing transactions, The batch size equals the number of packets the driver polls from the NIC in a single `recv()` operation (at most 32).

**NFork profiler.** For each invocation of the built-in data structure operations, the profiler keeps the arguments, the data structure's internal state, and the thread ID. Upon transaction aborts, the profiler records the conflicting data structure operations. In our current implementation, after the NF finishes execution, the profiler invokes a function provided by the corresponding built-in data structure to map each abort into its conflict cause. To prevent results from being skewed by profiling, we minimize the profiler overhead with several optimizations. Most importantly, since NFork's transactional memory is based on object multi-versioning [29], the profiler records the timestamps of the conflicting data structure operations in the critical path and then obtains the data structures' internal state off the critical path.

### 4.8 Developing concurrent NFs with NFork

We use a DDoS attack detector NF (Anti-DDoS) adapted from NFF-Go [2] to illustrate the process of developing concurrent NFs with NFork. Anti-DDoS checks whether >=80% of the incoming TCP flows have only one packet, in order to detect SYN flooding and similar attacks.

A packet set in Anti-DDoS consists of all TCP packets with the same source and destination IP and port (i.e., each TCP flow); the definition is shown in Listing 1.

Listing 2 shows the code of Anti-DDoS. The aggregate state is a distributed object struct (`ctrs`) with two members (Line 4); `flows` records the total number of flows, while `onePktFlow` records the number of flows with only one packet. The `init_pkt_set_handler` (Line 13) handles the first packet of a TCP flow. It increments `flows`, temporarily marks the flow as a one-packet flow by incrementing the `onePktFlow`, and then checks for a DDoS attack. Upon receiving subsequent packets of a flow, NFork invokes the `pkt_handler` (Line 26) to decrement `onePktFlow`. Finally, when a flow expires, the `expired_pkt_set_handler` (Line

```
1  #define CTR_STALENESS_MS 0 // flow counter staleness in ms
2  /* NF state definitons */
3  pkt_set_state bool isOnePktFlow;
4  aggregate_state distri_obj_t (CTR_STALENESS_MS) ctrs {
5    int flows, onePktFlows;
6  };
7  /* Callbacks for merging flow counter partitions */
8  void ctrs_merge(merged, partition) {
9    merged->flows += partition->flows;
10   merged->onePktFlows += partition->onePktFlows;
11 }
12 /* Handle first packet of a flow */
13 int init_pkt_set_handler(pkt, incoming_dev) {
14   isOnePktFlow = true;
15   distributed_obj_update(ctrs, flows++, onePktFlows++);
16   // Register the packet set to finish init
17   register_pkt_set();
18   // read flow counters and detect DDoS
19   ctrs_l = distributed_obj_read(ctrs, ctrs_merge);
20   if (ctrs_l->onePktFlows > 0.8 * ctrs_l->flows)
21     handle_ddos();
22   else
23     send_pkt(pkt);
24 }
25 /* Handle subsequent packets of a flow */
26 int pkt_handler(pkt, incoming_dev) {
27   if (isOnePktFlow) {
28     isOnePktFlow = false;
29     distributed_obj_update(ctrs, onePktFlows--);
30   }
31   send_pkt(pkt);
32 }
33 int expired_pkt_set_handler() {
34   distributed_obj_update(ctrs, flows--);
35   if (isOnePktFlow)
36     distributed_obj_update(ctrs, onePktFlows--);
37 }
```
Listing 2: Anti-DDoS written with NFork.

33) decrements `flows` and then checks if the flow had only one packet and, if so, decrements `onePktFlow`.

With the default semantics, Anti-DDoS always reads the latest value of the flow counters when detecting DDoS attacks. Under a simulated attack traffic with many small flows, it does not scale well with this default semantics (green series in Figure 4(a)).

To improve scalability, the developer uses the NFork profiler to profile Anti-DDoS and follow the recipes. The profilier's output, presented in Figure 4(b), shows that 97.5% of the transactions abort. 100% of the aborts are due to the first conflict cause of the concurrent read and update operations of distributed object `ctrs`. The profiler further provides the recipe: increasing the maximum allowed staleness of the distributed object (Table 5). The developer applies this recipe by increasing `CTR_STALENESS_MS` on Line 1 to 0.1ms. The outcome is an acceptable delay in detecting a DDoS attack. As shown in Figure 4(a), Anti-DDoS scales well after this semantic relaxation, increasing the throughput by 18× at 23 cores (from 4.5 to 79 mpps).

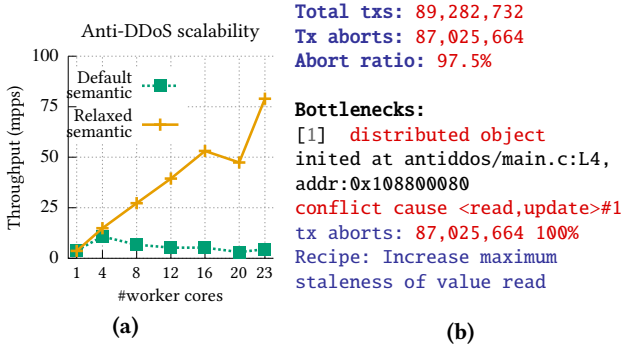

**Figure 4:** **(a)** Anti-DDoS throughput (max 0.1% packet loss) with its default semantics (read latest flow counter value) vs. relaxed semantics (allowing for stale read with a maximum staleness of 0.1ms). **(b)** Profile of Anti-DDoS with its default semantics.

## 5 Evaluation

We evaluate NFork by answering the following questions:

- Can NFork be used to develop a wide range of NFs? Does it hide concurrency from developers? (§5.2)
- What is the scalability of NFs written with NFork and how do they compare to hand-parallelized NFs? (§5.3)
- Can developers productively improve the scalability of NFs using NFork's profiler and recipes? (§5.4)

### 5.1 Evaluation setup

We port to NFork three NFs from Cisco's VPP [16] and compare them with their original hand-parallelized implementations. They are a MAC-learning bridge (Bridge), a load balancer (LB), and a NAT. Bridge [3] forwards packets to a destination network based on the packet's MAC address and learns MAC-to-network mappings from each packet. LB implements the Maglev [14] algorithm. NAT is endpoint-independent [26, 50] and translates a LAN endpoint's IP and TCP/UDP port to the same public value independent of the connecting WAN endpoint. We further implement with NFork a stateful firewall (FW) that blocks external connections and a traffic policer (Pol) that rate-limits traffic by destination IP.

Our servers have two sockets equipped with 24-core Intel Xeon 6248R processors, 256 GB DRAM, and Intel E810 100Gbps NICs. All the NFs use DPDK 20.11, and we tune them for the best performance following DPDK's official suggestions [22]. We use MoonGen [15] as the traffic generator, which transmits up to 98 million packets per second (mpps) per NIC port. The generator sends packets to the NFs on a different server and receives the processed packets from the NFs.

We stress the NFs using real Internet traffic and (for Bridge) a university data-center synthetic trace. For the former, we use the CAIDA trace [17]. Its average TCP/UDP flow size is 21 packets. For NAT and FW, we simulate LAN-to-WAN traffic with the CAIDA trace. For LB, we modify the CAIDA trace to keep only the HTTP(s) packets and rewrite the des-tination IP of packets to LB's virtual IP. For Bridge, we use a synthetic trace simulating two networks connecting through it. The trace generates a flow with 10,000 packets for each pair of MACs and each direction, the same as a university data center's average MAC flow size [51].

| NF | Packet set | Packet-set state | Aggregate state |
|---|---|---|---|
| Endpoint independent NAT [26, 50] | none | none | Public <IP:port> allocator; address mappings |
| Load balancer [14] | Incoming TCP flows | Flow-server mapping | Server health status |
| Bridge [3] | none | none | MAC-port mappings |
| DNStunnel detector [7] | none | none | Orphan DNS response counters |
| Sidejacking detector [7] | TCP session | Session context | Client IPs of each HTTP session |
| DDoS detector [2] | TCP/UDP flow | flow size | flow statistic |
| PRADS [58] | TCP/UDP session | Session context | Host info; traffic statistic |
| Web cache [41] | TCP session | Session context | Cached responses |
| Stateful firewall [61] | TCP/UDP session | Session firewall policy | none |
| Policer [38] | Packets to a LAN host | Incoming packet rate of the host | none |
| Portscan detector [47] | Packets from a host | Maliciousness probability | none |
| IPSec tunnel [45] | Tunnel session | Encryption key | none |
| nDPI [25] | TCP/UDP session | Application protocol | none |
| Abacus [25] | TCP session | Data buffer | none |

**Table 6:** Popular NFs can be easily developed with NFork.

### 5.2 Generality and Concurrency Masking

**Generality.** We use 14 NFs widely used in industry and academia to assess the generality of NFork's programming model (Table 6). We analyze the code and/or the semantics of these NFs "by hand" and conclude that NFork's programming model can express all the NFs. Furthermore, defining packet sets is straightforward, with most of them being either flows or sessions.

**Simplifying concurrent NF development with concurrency masking.** We port to NFork 3 concurrent NFs (Bridge, LB, NAT) from the VPP family (§5.1). We use them to study whether NFork can mask concurrency from developers, thereby simplifying concurrent NF development. The concurrency patterns in the VPP NFs are (i) state partitioning among cores (flow-server mappings in LB, address mappings, and public <IP:port> pairs in NAT) and (ii) concurrent data structures (lock-free MAC table in Bridge, server table in LB, and lock-free queues in NAT).

For all the NFs we port, NFork frees the developers from handling the concurrency patterns discussed above. Instead,

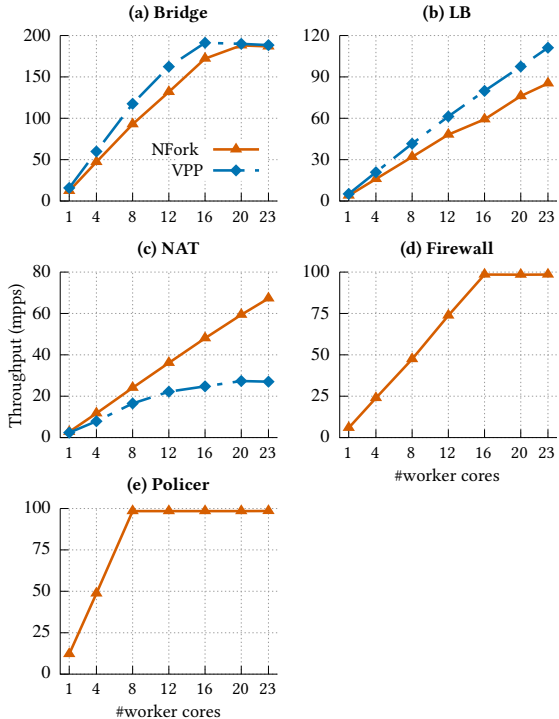

**Figure 5:** Throughput of the evaluated NFs with < 0.1% packet loss. The maximal number of worker cores is 23 since both VPP and NFork dedicate one core for extra tasks such as monitoring.

developers only need to add, in total, 35 – 58 LOC to specify packet sets (0 – 5 LOC), categorize state (7 – 15 LOC), use the built-in data structures (21 – 40 LOC), and structure the handlers (7 – 16 LOC). The added code represents, at most, a few percentage points of the LOC implementing the NF logic.

**Summary.** We show that the NFork programming model is sufficiently general to express the semantics of widely used NFs, is sufficiently familiar to NF developers to be used productively (e.g., packet sets are easy to define), and can fully abstract away the challenges of concurrent code.

### 5.3 NF Scalability, Overhead, and Latency

Figure 5 shows the throughput of the NFs written in NFork, compared to VPP NFs. We ensure fair comparison by making the semantics of NFork NFs and VPP NFs the same. Specifically, the MAC table entry in Bridge uses a refreshing interval of 1 minute, and NAT uses 57 public IPs. With our traces and NF setup, NAT and Bridge update aggregate state every 10 and 5,000 packets, respectively, while the other NFs have rare or no updates to aggregate state.

In summary, NFork NFs scale linearly, and 3 out of the 5 reach the limit of the traffic generator or the servers (98 mpps for the firewall and policer using one NIC port and 188.8 mpps for Bridge using two NIC ports). Furthermore, NFork NFs scale similarly to or better than the hand-parallelized NFs in VPP. NFork has, on average, 20% and 23% lower throughput than VPP with Bridge and LB, respectively, while outperforming VPP in NAT by up to 2.5×.

For Bridge, before 20 cores, NFork is slower than VPP due to the overhead incurred by the transactional memory, as detailed later. For LB, similar to Bridge, almost half of the overhead in NFork is incurred by the transactional memory. Even so, as we'll see shortly, latency is competitive with the VPP equivalents. The rest of the LB's overhead is due to the higher cache miss rate on accessing flow-to-server mappings, which we confirm using perf. Specifically, the NFork LB stores the <src/dst IP/port, L4 proto> 5-tuple as the key. Meanwhile, the VPP LB uses a clever and specific optimization to store a 32-bit hash for each flow, which saves nine bytes per mapping compared to the NFork LB. Given the 5.8 million table entries, such saving results in a higher cache hit rate.

We have explained why VPP NAT cannot scale in §2.3: it suffers from contention on the shared queues for forwarding packets of a LAN endpoint to the same core, which is necessary for partitioning address mappings among cores. Meanwhile, the NFork NAT does not partition address mappings and scales well since, contradicting VPP developers' assumptions, the contention on the address mappings is actually low, at least for the workloads we use.

**Microbenchmarks.** We use microbenchmarks to study how NFork behaves under different workloads. We develop a dummy NF in NFork, whose aggregate state is an array with 10,000 elements. We simulate different workloads by varying the access types (**RO**: read-only, **RW**: read and write per packet) and skewness (Zipfian values of 0 and 0.99) to the array.

Figure 6(a) shows the result. As expected, NFork scales linearly with read-only workloads and reaches the limit of the traffic generator after 20 cores. This is because concurrent reads do not abort transactions. For the write workload, NFork scales well as long as the contention on the aggregate state is relatively low. The worst case for NFork is the write workload with high skewness to the aggregate state. Our experience with real-world NFs suggests that this situation is uncommon. Furthermore, in such a case, as detailed in §5.4, NFork can guide developers to identify opportunities for improving scalability through semantic relaxation.

**Handler batching.** As mentioned in §4.7, NFork batches the handlers of multiple packets in the same transaction to amortize the overhead of transaction start and commit. Figure 6(b) shows the average speedup of batching, varying the number of worker cores from 4 to 23. On average, batching improves the NF throughput by 34%.

**Cost of using transactional memory.** We analyze the cost of using transactional memory to abstract away concurrency. Figure 6(c) shows the overhead reported by perf. The transaction overheads are nearly constant, explaining the linear scalability of the evaluated NFs. Bridge's transaction overhead reduces after 12 cores because it reaches the limit of the testing server and is thus underloaded. The transaction does not add overhead to the firewall and policer since they do not have aggregate state. The synchronization overhead in hand-optimized concurrent NFs may be better or (a lot) worse, as shown in Figure 5.

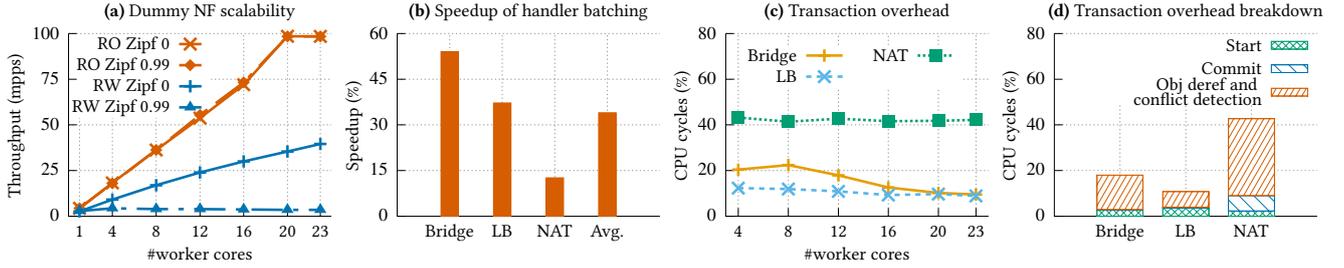

**Figure 6: (a)** Throughput (max 0.1% packet loss) of a dummy NF with varying aggregate state access patterns. **(b)** Average NF performance speedup with batching. **(c)** Transactional memory overhead (measured as the percentage of CPU cycles spent in transaction runtime). **(d)** Breakdown of transactional memory overhead.

Figure 6(d) shows the breakdown of the transaction overheads with 12 cores. Since NFork's transactional memory is based on multi-versioning, most of the overhead comes from object dereferencing (i.e., obtaining the right version of objects) and detecting conflict accesses.

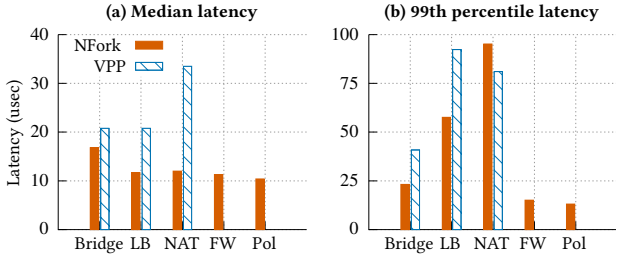

**Figure 7:** End-to-end packet latency of NFs.

**Latency.** Figure 7 shows the end-to-end packet latency (i.e., the latency between when the NF receives a packet on the NIC and when it sends the packet on the wire) of the evaluated NFs with 23 worker cores. The evaluated NFs operate with their maximal throughputs with less than 0.1% packet loss. VPP NFs have higher latencies in general due to aggressive packet batching.

**Memory overhead.** We report the memory footprint of NFork NFs by measuring the maximum resident set size during execution. We find that the memory footprints of NFork NFs increase only slightly with the number of cores increasing from 1 to 23 with a maximal increment of 19% (Bridge: 39.5 to 47.1 MB; LB: 27 to 28.4 MB; NAT: 3,713.9 to 3,726.2 MB; Firewall: 25.5 to 26.1 MB; Policer: 24.8 to 25.6 MB). This slight increase comes from the per-core logs used in the NFork transactional object framework.

**Summary.** We conclude that the throughput and latency of NFs developed with NFork is on par with production-grade hand-parallelized NFs, so the improved developer productivity and avoidance of concurrency bugs (§5.2) offered by NFork do not come at the price of severely reduced performance.

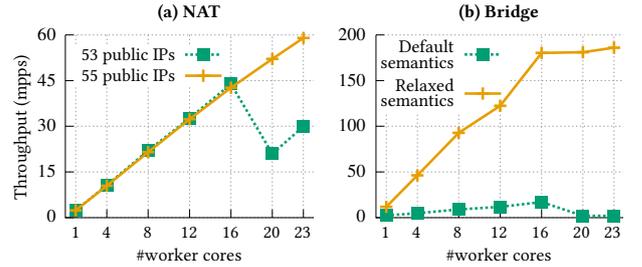

**Figure 8: (a)** NAT throughput (max 0.1% packet loss) with different number of public IPs. **(b)** Bridge throughput (max 0.1% packet loss) with the default semantics (MAC table entry refreshed on each packet) vs. the relaxed semantics (refresh interval 1 second).

### 5.4 Improving NF Scalability with NFork

Subsection 4.8 already presents one case study of improving NF scalability with NFork. This subsection presents two more case studies with NAT and Bridge. Each MAC table entry (i.e., MAC-to-network mapping) in Bridge has a validity duration of 2 minutes, after which Bridge deletes the entry. Bridge can refresh a MAC table entry to reset its validity duration.

We port to NFork a sequential implementation of NAT and Bridge. We refer to the semantics in the sequential implementation as the *default semantics*, where NAT uses 53 public IPs and Bridge removes a MAC table entry only if it has not received a packet from the corresponding MAC for more than the validity duration. While such semantics perform well in the sequential implementation, as detailed later, they make NFs inherently difficult to scale.

**NAT.** Figure 8(a) shows that NAT with the default semantics stops scaling after 16 cores. We profile NAT with the NFork profiler at 23 cores. The profiler shows that 99.4% of the transaction aborts are due to concurrent allocations (Table 3) made by an allocator that manages free public <IP:port> pairs. The profiler also provides the recipe to overprovision the number of public <IP:port> pairs. We follow the recipe, and as shown in Figure 8(a), NAT scales linearly with 55 public IPs, increasing throughput by 2× at 23 cores. Although public IPs are precious resources, under performance-critical scenarios, we believe that overprovisioning by a few percent will be worthwhile given the significant scalability improvement.

**Bridge.** Figure 8(b) shows that Bridge cannot scale with the default semantics. To maintain this semantics, Bridge has to refresh a MAC table entry each time it receives a packet from the corresponding MAC address. The NFork profiler shows that with 23 cores, 63.5% of the transactions abort. Furthermore, almost all the aborts are due to concurrent refreshing of MAC table entries (Table 3). We follow the NFork recipe and increase the refreshing interval to 1 second (i.e., Bridge will only refresh a MAC table entry if it has not been refreshed for more than one second). Figure 8(b) shows that this change makes Bridge scale linearly. At 23 cores, its throughput increases by $91\times$ (from 2.04 to 186 mpps).

Increasing the refreshing interval relaxes Bridge's semantics since the MAC table entries could be removed (up to 1 second) earlier than expected, potentially leading to packet flooding. This is acceptable given (i) the improvement in common-case scalability and (ii) the fact that packet flooding would be rare; the new refreshing interval (1 second) is still much smaller than the validity duration (2 minutes).

**Summary.** We show that NFork aids developers to productively improve scalability for NFs that are inherently difficult to scale. The NFork profiler directly shows the root causes of the scalability bottlenecks, and the developers only need to apply the corresponding recipe. Applying recipes requires changing only a few lines of NF code (one line for both NAT and Bridge). Developers can use their domain knowledge to make the trade-off between scalability and semantics.

## 6 Limitations

**Packet set definition.** The packet set abstraction introduces a trade-off between load balancing and shared state contention. Fine-grained packet sets (i.e., those usually consist of a small number of packets) can achieve good load balancing among cores but may introduce a large aggregate state. This, in turn, may increase the contention on the aggregate state and thus limit scalability. Coarser-grained packet sets minimize the aggregate state but may incur load imbalance. This is because the number of packets in each packet set may differ. Hence, a few large packet sets could saturate a core while the other cores are left idle. Such a load imbalance may again limit performance. The NFork profiler could aid developers in making this trade-off by reporting the load imbalance among cores. For the NFs we study (§5.2), we find that defining a packet set is straightforward and do not expect the load imbalance problem with our packet set definition.

**Starvation freedom.** Our current implementation does not guarantee starvation freedom. A thread could starve if it accesses highly-contended aggregate state objects. Fortunately, this is usually not an issue for NFs. For the NFs we study (§5.2), most of them have low contention on aggregate state objects. For the rest (e.g., Anti-DDoS, the Bridge, and the NAT), we can minimize the contention through semantic relaxations.

**Load balancing.** Our current implementation uses RSS to achieve load balancing. Our evaluation with the real CAIDA trace does not show load imbalance issues, but they are theoretically possible. In the future, we plan to integrate more advanced mechanisms like RSS++ [5] in NFork.

## 7 Conclusion

We present NFork, a system for NF domain experts to *productively* develop, profile, and optimize concurrent NFs. NFork enables developers to write NFs as sequential programs with the *packet set* abstraction. This simplifies NF development and eliminates concurrency problems at the source. Leveraging NF characteristics, the NFork runtime parallelizes the processing of packets with *transactional memory* and *built-in data structures*, thereby achieving scalability while eliminating concurrency bugs. The NFork profiler bridges the semantic gap between low-level events and application-level concurrency and, by using the conflict cause abstraction, it reveals to developers the root causes of scalability bottlenecks. When these are inherent to the NF's semantics, the concrete recipes help developers relax semantics in exchange for scalability. Our experiments show that NFs written in NFork have similar or better scalability with manually parallelized VPP NFs [16], and developers can effectively improve NF scalability following the NFork profiler and recipes.